\documentclass[manuscript]{aastex}

\slugcomment{}

\shorttitle{Alignment effect of galaxies}
\shortauthors{Hung et al.}

\begin{document}


\title{The galaxy alignment effect in Abell 1689:
evolution, radial and luminosity dependence}


\author{Li-Wei Hung\altaffilmark{1}, Eduardo Ba{\~n}ados\altaffilmark{2},
Roberto De Propris\altaffilmark{3}, Michael J. West\altaffilmark{4}}

\altaffiltext{1}{Department of Astronomy, Ohio State University, Columbus, OH, USA}
\altaffiltext{2}{Departamento de Astronomia y Astrofisica, Pontificia Universidad
Catolica, Santiago, Chile}
\altaffiltext{3}{Cerro Tololo Inter-American Observatory, La Serena, Chile}
\altaffiltext{4}{European Southern Observatory, Santiago, Chile}



\begin{abstract}

We measure alignments on scales of 1 Mpc $h^{-1}_{71}$ for galaxies in
Abell 1689 ($z=0.18$) from an existing Hubble Space Telescope mosaic. We 
find evidence of galaxy alignment in the inner 500 $h^{-1}_{71}$
kpc. The alignment appears to be stronger towards the centre and is
mostly present among the fainter galaxies, while bright galaxies are
unaligned. This is consistent with a model where alignments originate 
from tidal locking.

\end{abstract}





\section{Introduction}

Alignments between galaxies and their host structures is predicted to occur
in $\Lambda$CDM models (e.g., \citealt{faltenbacher05}; Hopkins, Bahcall \&
Bode 2005; \citealt{basilakos06}). For instance, in turbulence models, galaxies'
major axes should align with the major axis of the closest large scale structure
\citep{shandarin74,efstathiou83}, while Navarro, Abadi \& Steinmetz (2004) relate
the alignment of spiral galaxies to the acquisition of angular momentum via tidal
torques.

Perhaps the best known example of preferential galaxy alignments is the 
observation that the major axis of brightest cluster galaxies (BCGs) is oriented 
along the distribution of cluster members and `points' towards other nearby 
clusters on scales of $\sim 10$ -- $20$ Mpc h$^{-1}$ (e.g., \citealt{binggeli82,struble90};
Trevese, Cirimele \& Flin 1992; Fuller, West \& Bridges 1999; \citealt{niederste10}
among others). This is explained by the formation of BCGs via a process of collimated
infall along filaments feeding the cluster growth \citep{west94,dubinski98}.

It is unclear whether other cluster galaxies other than the BCGs are expected to 
show preferential alignments (direct alignments with the BCG axis) in CDM models
(e.g., West, Villumsen \& Dekel 1991). Any primordial alignments may not survive
the cluster environment, as they are erased by strong dynamical interactions \citep{
coutts96,plionis03}. On the other hand, the  cluster tidal field may eventually 
induce new alignments by a process of tidal locking akin to the Earth-Moon system 
(\citealt{ciotti94}, Pereira, Bryan \& Gill 2008).

Direct galaxy alignments have been detected for some clusters \citep{plionis03,aryal06}
but not in others (\citealt{strazzullo05}; Aryal, Kandel \& Saurer 2006). It is
likely that the alignments depend on mass, cluster age and position in complex
ways. Arguably, the best studied example is the Coma cluster. In a comprehensive study 
using high quality multi-color imaging and spectroscopy, Torlina, De Propris \& West (2007) 
found no evidence of preferential alignments other than for the two brightest cluster 
members for the inner 400 $h^{-1}$ kpc of the Coma cluster. \cite{adami09} broadly confirm 
this result but find some evidence of alignment for the fainter cluster members and in areas 
that appear to coincide with substructure. \cite{plionis03} detect a clear alignment signal 
in Abell 521, but this is a dynamically young object containing rich substructures 
and the alignment appears to correlate with the individual subclusters.

The evolution of the alignment effect offers clues to its origin and possibly
constraints on galaxy formation models and their interaction with large scale
structure. We may expect that higher redshift clusters, observed closer to the
epoch of their formation, may yield a stronger signal if primordial alignments
are important. In the models of \cite{pereira08} initial alignments are quickly
destroyed, but the direct alignment increases with time because of tidal torques.
\cite{niederste10} have studied the alignment of BCGs with their clusters at $0
< z < 0.4$ and found that more dominant central galaxies are more strongly aligned 
and that the effect is stronger at lower redshifts, a result that can be interpreted 
within the framework of hierarchical accretion and growth.

In this paper we study the alignment effect in Abell 1689 over a $\sim 1$ Mpc $h^{-1}$
field from a wide mosaic of Hubble Space Telescope (HST) data. This is the most distant
cluster where the alignment effect has been searched for. HST observations provide the
necessary spatial resolution and stable point spread function necessary to measure the
position angles and ellipticities of faint galaxies at high redshift. We describe the
data and analysis in section 2. The results are reported in section 3 and discussed in
section 4. We adopt the latest WMAP7 cosmological parameters: $\Omega_M=0.27$, $\Omega_{\Lambda}
=0.73$ and H$_0=71$ km s$^{-1}$ Mpc$^{-1}$.

\section{Data and Analysis}

The data used in this project consist of a $4 \times 4$ mosaic of WFPC2 (Wide Field
and Planetary Camera 2) images covering a total of $10'$ on the side (PID: 5993;
PI: Kaiser). Exposure times were 1800s in $V$ (F606W) and 2300s in $I$ (F814W). All
data were retrieved as fully processed images from the Hubble Legacy Archive website.

Photometry for objects in these fields was carried out using Sextractor \citep{bertin96}.
Further details about the photometry may be found in a companion paper by Ba{\~n}ados
et al. (2010, submitted) dealing with the galaxy luminosity function in Abell 1689, but 
we summarize the most relevant points here. Sextractor was run twice, once with parameters
appropriate to the detection and photometry of bright galaxies, without excessively 
deblending their images, and then with parameters suited to faint galaxies. The same
parameters as in \cite{leauthaud07} were used. This allowed us (see Ba{\~n}ados et al.
for details) to use the COSMOS counts to statistically subtract the fore/background
contribution to determine the cluster LF.
 
We selected galaxies in the $I$ band, both because this is the COSMOS selection band and 
because this is better related to the stellar mass. However, we also measure two aperture 
magnitudes in $V$ and $I$, to derive galaxy colors and use these to identify a sample of 
cluster members via their position with respect to the cluster red sequence (see below). 
We also measure, for each galaxy, its ellipticity and position angle. We use the values
determined from the $I$ band, but we find that the values from the $V$ band images are
identical to within a few \% (Figure 1). All images were visually inspected to remove 
spurious objects, satellite trails and (especially) arclets. Magnitudes were calibrated 
on to the AB system with tabulated zeropoints.

In order to use these data for the alignment effect we need to isolate a sample of
cluster members. Lacking a large redshift survey or numerous bands to measure
photometric redshifts, we use galaxies on the color-magnitude relation as these
are likely to consist predominantly of cluster members. We plot the $V-I$ vs. $I$
color-magnitude diagram of galaxies in Abell 1689 in Figure 2, where we assume
that galaxies within $\pm 0.3$ mag. of the color-magnitude defined by the brighter
galaxies are cluster members. The faint-end limit of our selection is set by
our ability to carry out star-galaxy separation (see Ba{\~n}ados et al. 2010,
their Figure 1). Note that the red sequence LF in Ba{\~n}ados et al. is in
very good agreement with their total LF, arguing that most cluster members
indeed lie on the red sequence. However, this selection limits the sample
of objects to E/S0 galaxies and dwarf `ellipticals'; some galaxies in the blue
cloud may be cluster members, but we have no way of determining membership,
other than by the red sequence, and including them in our study, with present 
data.

In Coma, \cite{adami09b} find that the scatter on the red sequence increases
for $M_R > -18$, while in Virgo \citep{janz08} find an increase in the scatter
at $M_R > -14$; we instead use a rectangular aperture around the best fit
to the color-magnitude relation to select members. However, we choose to be 
conservative in our selection, as enlarging the color 'box' used in Figure 2 
runs the risk of including an increasing fraction of non-cluster members, 
especially at the faint end where the field counts are increasing steeply, and
where color errors become more significant (preferentially scattering blue galaxies
on to the red sequence), and therefore potentially diluting the sample. This might 
indeed be an issue in the cluster outskirts, where the relative fraction of cluster 
members decreases rapidly.

As shown by \cite{holden09} we need to be careful to understand the limits
of our method to obtain reliable measurements of the galaxy position angle.
This depends on the spatial resolution of the image, the point spread function
and the ellipticity of galaxies (round objects have no real orientation, of
course). In order to do this, we carry out a series of simulations, by placing
artificial images with $17 < I < 25$ (the approximate star/galaxy separation
limits in Banados et al. 2010), $0.0 < e < 0.7$ and random position angles, using 
the IRAF/ARTDATA package. We proceed to detect these galaxies in the same fashion 
as our targets and measure how the ellipticity and position angle are recovered 
as a function of input ellipticity and apparent magnitudes. All the images were 
convolved with a Gaussian having appropriate FWHM for HST images. We used the 
size-luminosity relation for Virgo galaxies by \cite{janz08} to derive input sizes 
as a function of magnitude, using de Vaucoleurs profiles for galaxies brighter than 
$M_I=-18$ and exponential profiles for fainter ones. A total of 100 galaxies per 
magnitude (in 1 mag. interval) and ellipticity (in 0.05 intervals) steps were simulated, 
for a total of 14,000 objects. 

The results of this exercise are shown in Figures 3 and 4, plotting $\Delta e$ (input 
minus output) vs. $I$ as a function of input $e$ and $\Delta$ PA vs. $I$, also as a 
function of input $e$. Based on these plots, we choose to trust only galaxies for which 
we can determine $e$ to within 0.1 and the PA to $20^{\circ}$. Targets for our study are
therefore galaxies with $I < 24$, $e > 0.2$ and lying on the red sequence in Figure 2.

\section{The alignment effect in Abell 1689}

Figure 5 (upper panel) shows the PAs of all galaxies used across the cluster
field, plotted as a segment. The brightest cluster galaxy is identified as
a thick red segment. The middle panel shows a histogram of the distribution
of position angles, while the bottom panel presents the results of a Kuiper
test, to determine whether there is alignment and its significance (cf.,
\citealt{torlina07}).

According to this test, there is a marginal $\sim 1.5\sigma$ detection of
alignment in this cluster. If alignment exists, we are also interested in
whether this is primordial or newly induced by tidal interactions. In the
models by \cite{pereira08} the alignment signal should rise towards the
cluster centre but drop again closest to the BCC, while \cite{aragon07}
detected a weak primordial alignment signal by stacking several hundred
clusters from the SDSS (in comparison, \citealt{torlina07} find no alignment
along the filaments surrounding the Coma cluster).

We plot the histogram of PAs and the Kuiper test results for regions within 
0 -- 200 kpc (projected) from the cluster centre, 200 -- 400 kpc, 400 -- 600 
kpc and 600 -- 1000 kpc, in Figure 6 (with cosmology as indicated above). The 
results are that there is relatively significant alignment ($P$ values of about 
0.12 or about $2\sigma$) in the two inner regions, but no detection in the two 
outer ones. In addition, the PA of the maximum deviation from uniformity is $\sim 
30^{\circ}$ -- $60^{\circ}$ in both the two inner regions and this is also very 
close to the measured PA of the brighter cluster galaxy ($26^{\circ}$), arguing 
that the measured alignment is likely to be real and to consist of direct alignment
between galaxies and the BCG. However, the lack of alignment signal in the
outer regions may at least partly be due to the increased degree of contamination
from field galaxies.

As a test, we also consider the distribution of the acute angles between
the galaxy PAs and the great circle connecting each galaxy to the cluster
centre and find no evidence of a radial alignment (with galaxies pointing
to the cluster centre as in \citealt{pereira05}). We therefore conclude
that we find evidence of so-called direct alignment where galaxies point 
along the major axis of the cluster distribution and the BCG.

We may further ask what galaxies are responsible for the observed signal. In
Figure 7 we plot the histogram of PAs and Kuiper test for bright ($M_I < -18$)
and faint ($M_I > -18$) galaxies. The bright galaxies are not aligned, in 
good agreement with the Coma data of \cite{torlina07}, but the faint galaxies
show significant (better than 2$\sigma$) alignment, with PA consistent with
that measured in Figure 5. We infer from this that the alignment we measure is due
to a population of fainter galaxies residing prevalently in the cluster centre.

While the giant galaxies are not aligned, we may also ask whether there is an
aligned component. We can only classify galaxies into E or S0, because of our
color selection. For these objects, Figure 8 shows the results of the Kuiper
test. There is no evidence of alignment among the giant galaxies, irrespective
of morphology.

\section{Discussion}

Our results show evidence of an alignment signal in the inner region of Abell 1689
and also imply that while bright galaxies are not aligned, the fainter galaxies are 
responsible for most of the observed signal, at least in Abell 1689. The results for 
bright galaxies are in agreement with the work in Coma by \cite{torlina07}, while the 
stronger alignment for fainter galaxies is at least partially consistent with \cite{
adami09}.

The most comprehensive theoretical study of the alignment effect come from
\cite{pereira08}. Based on a series of N-body simulations in a $\Lambda$CDM
universe, \cite{pereira08} find that the alignment effect is independent of
cluster mass, that there is a degree of primordial alignment at large radii
and the alignment becomes stronger towards the cluster centre, owing to tidal
forces, while its strength should grow with decreasing redshift. The work by
\cite{plionis03} argues that the alignment is stronger for dynamically young
clusters, where galaxies still `preserve' the memory of their infall history.

Our results are in broad agreement with some of these predictions. We detect no
alignment in the outskirts, although the signal may be too weak for us to detect
\citep{aragon07}. In the cluster outskirts, contamination of the sample by non members 
would tend to dilute any alignment as well. In agreement with the expectations from 
theory, the alignment signal becomes stronger towards the cluster centre. The model 
by \cite{pereira08} makes no strong predictions concerning the size of the alignment 
effect as a function of satellite mass. In our case, the bright galaxies show no
significant signal, as they do in Coma \citep{torlina07}. This is true even when we
separate E from S0. Even in a sample of above 300 clusters at low redshift, \cite{pereira05} 
find only a weak signal. This is partly due to the difficulty of measuring the alignment 
of real stellar distributions as opposed to cleanly defined dark matter halos. However,
the alignment effect for giants tends to be much weaker than expected, even in younger
clusters. This suggest that there is either no primordial alignment, or that giants have
dwelled in cluster environments long enough for any original anisotropy to be erased,
while the processes that lead to alignments are either inefficient or have not had 
sufficient time to operate on the scales of luminous galaxies. 

Fainter galaxies are instead significantly aligned, especially in the centre.
This is consistent with the tidal model. For brighter galaxies, the change in 
position angle would take place slowly, over multiple orbits, and would be most 
effective for radial orbits \citep{pereira08}. Thus it is possibly not fully 
surprising if we detect no alignment signal. On the other hand, we might expect 
that fainter galaxies will be more strongly affected by the cluster tidal field 
and would tend to align themselves more rapidly, by analogy with the tidal locking 
process studied by \cite{ciotti94}.

It is clear that detection of galaxy alignments is difficult: it requires large
statistics, high resolution wide-field imaging to measure position angles reliably
and either redshift information or colors to select members. Nevertheless,
the alignment effect provides a useful test of models for galaxy formation
and dynamical evolution. The rich sample of clusters to be gathered by the
Multi-Cycle Treasury survey will provide us with a useful dataset to test
this issue further.

\acknowledgments
Based on observations made with the NASA/ESA Hubble Space Telescope, and obtained from the Hubble Legacy Archive, which is a collaboration between the Space Telescope Science Institute (STScI/NASA), the Space Telescope European Coordinating Facility (ST-ECF/ESA) and the Canadian Astronomy Data Centre (CADC/NRC/CSA).



{\it Facilities:} \facility{HST (WFPC2)}.

\clearpage

\begin{figure}
\epsscale{0.8}
\plotone{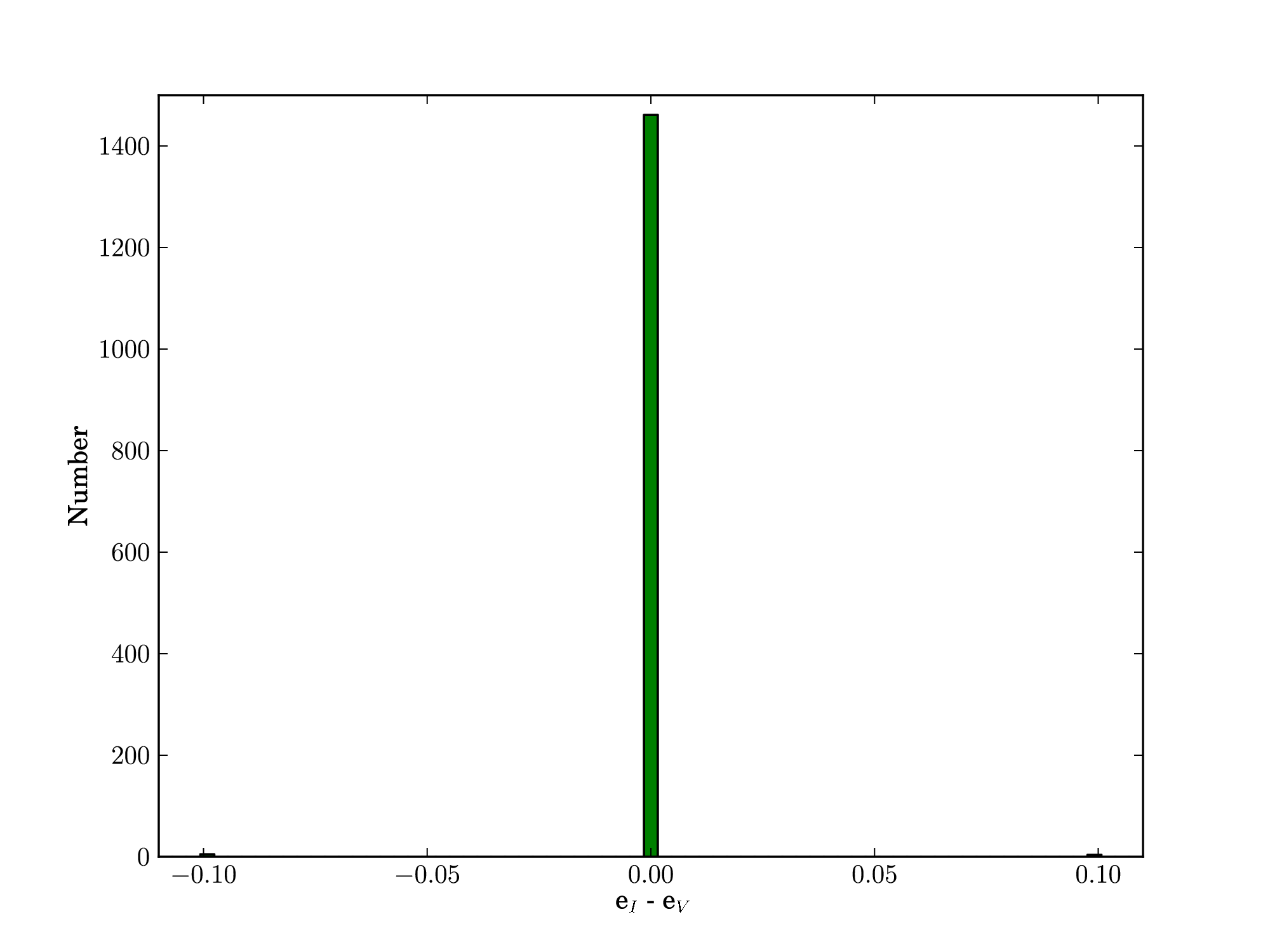}
\vspace{1cm}
\plotone{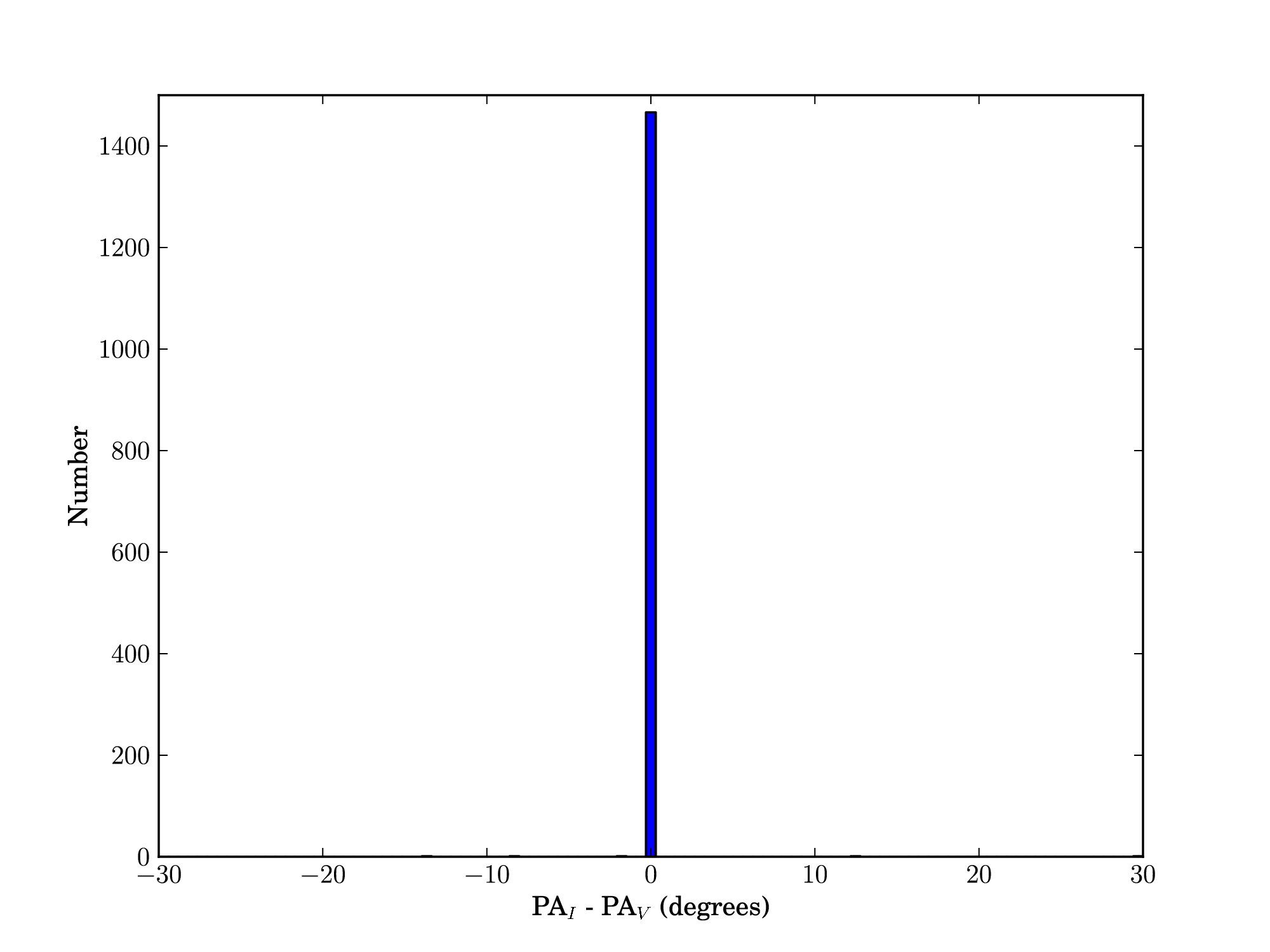}
\caption{Histograms of the difference between ellipticities
and position angles as determined from the $V$ and $I$ band
images. The mean is zero with a scatter of a few \% at most}
\end{figure}
\clearpage

\begin{figure}
\includegraphics[width=6in]{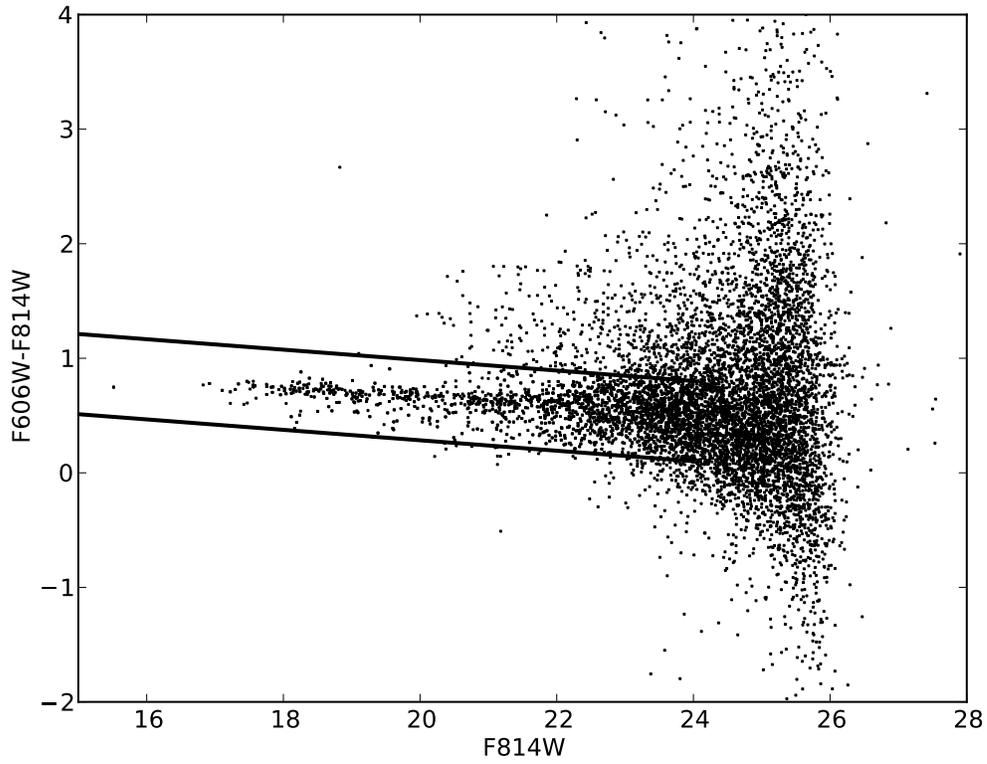}
\caption{Color-magnitude relations in $V-I$ for A1689. We 
overplot the selection range in color and magnitude that 
we adopt for membership in the cluster.}
\end{figure}
\clearpage

\begin{figure}
\plotone{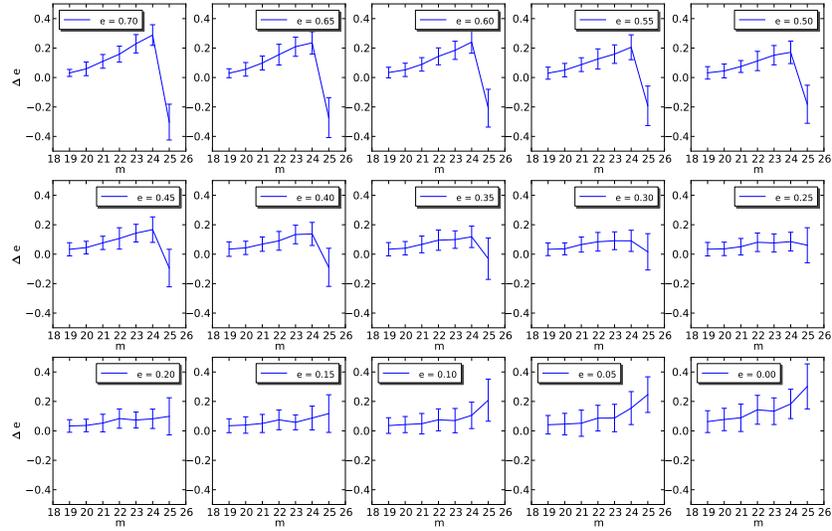}
\caption{Difference between input ellipticity and 
ellipticity output by Sextractor for simulated galaxies as a function 
of $I$ magnitude. Each panel corresponds to a different ellipticity 
identified in the legend.}
\end{figure}
\clearpage

\begin{figure}
\plotone{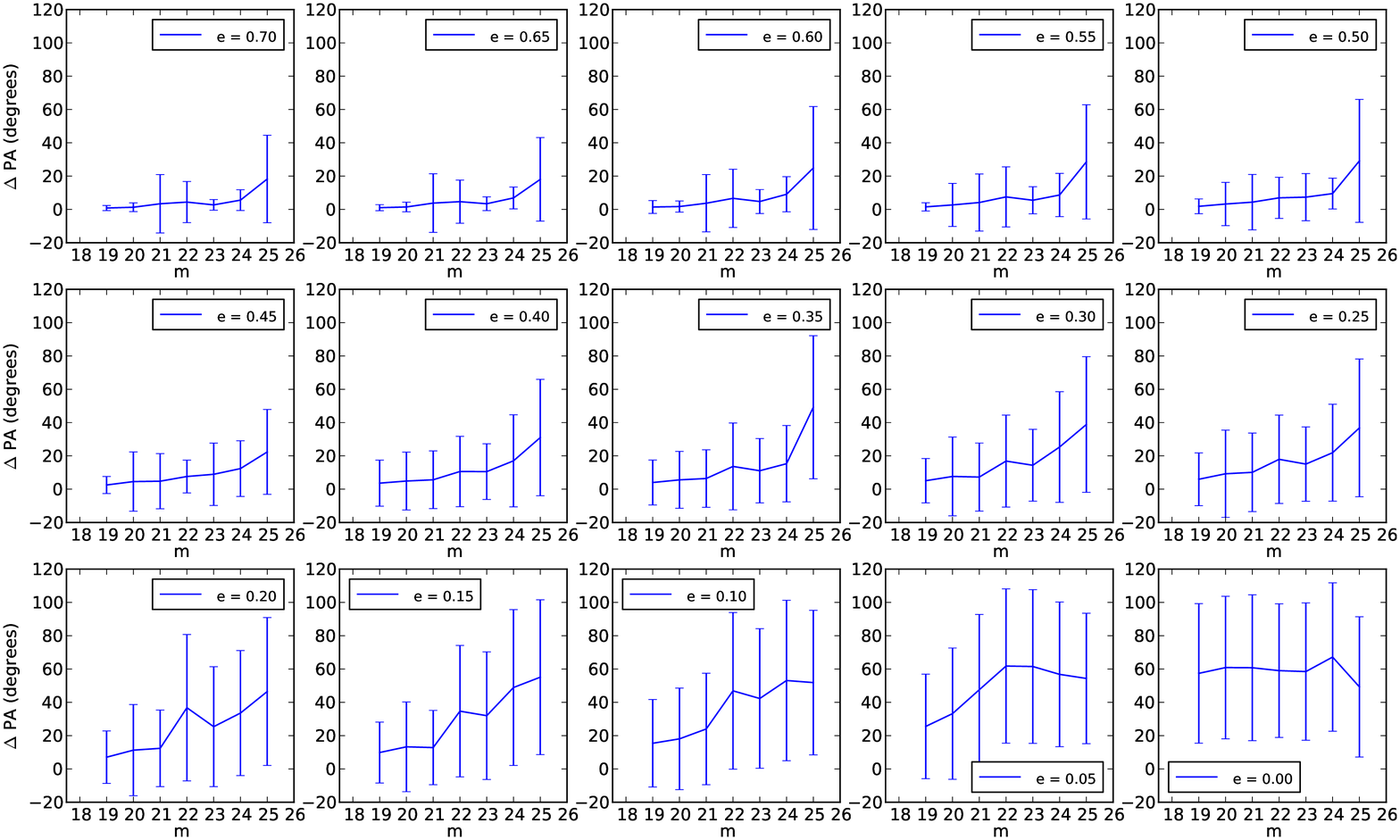}
\caption{Same as Figure 2 but for the difference between input position 
angle and PA output by Sextractor for simulated galaxies as a function of 
$I$ magnitude.}
\end{figure}
\clearpage

\begin{figure}
\includegraphics[width=4.5in]{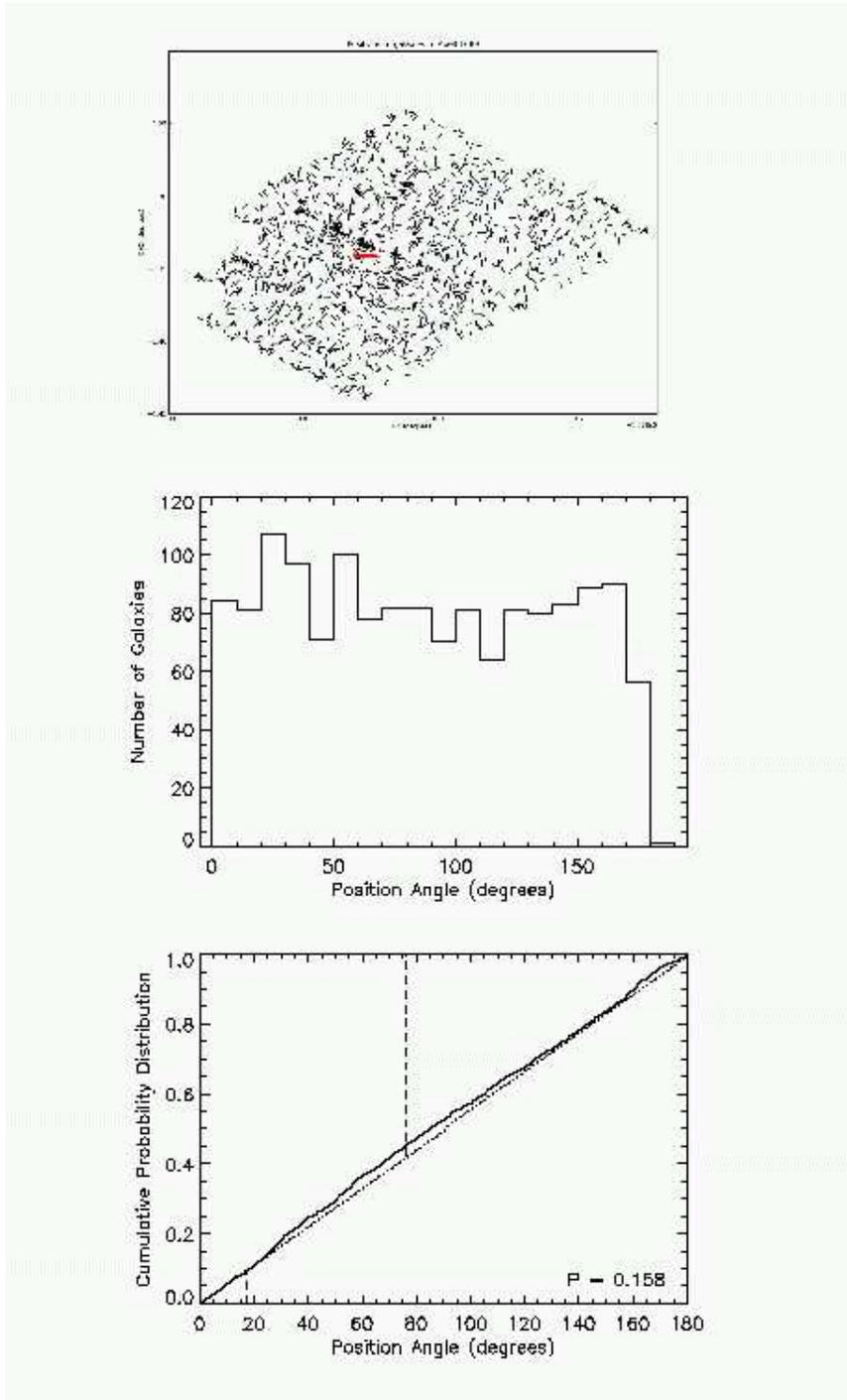}
\caption{\it Top panel: \rm Position angles of Abell 1689 galaxies, 
with the brightest cluster galaxy marked in red. \it Middle panel: \rm
Histogram of position angles. \it Bottom panel: \rm cumulative distribution
function compared to a uniform distribution. The P value returned by the
Kuiper test is indicated in the figure, as well as the angle of maximum
deviation. A lower P value indicates greater alignment. }
\end{figure}
\clearpage

\begin{figure}
\resizebox{6.5in}{7.5in}{
\includegraphics[140,70][580,720]{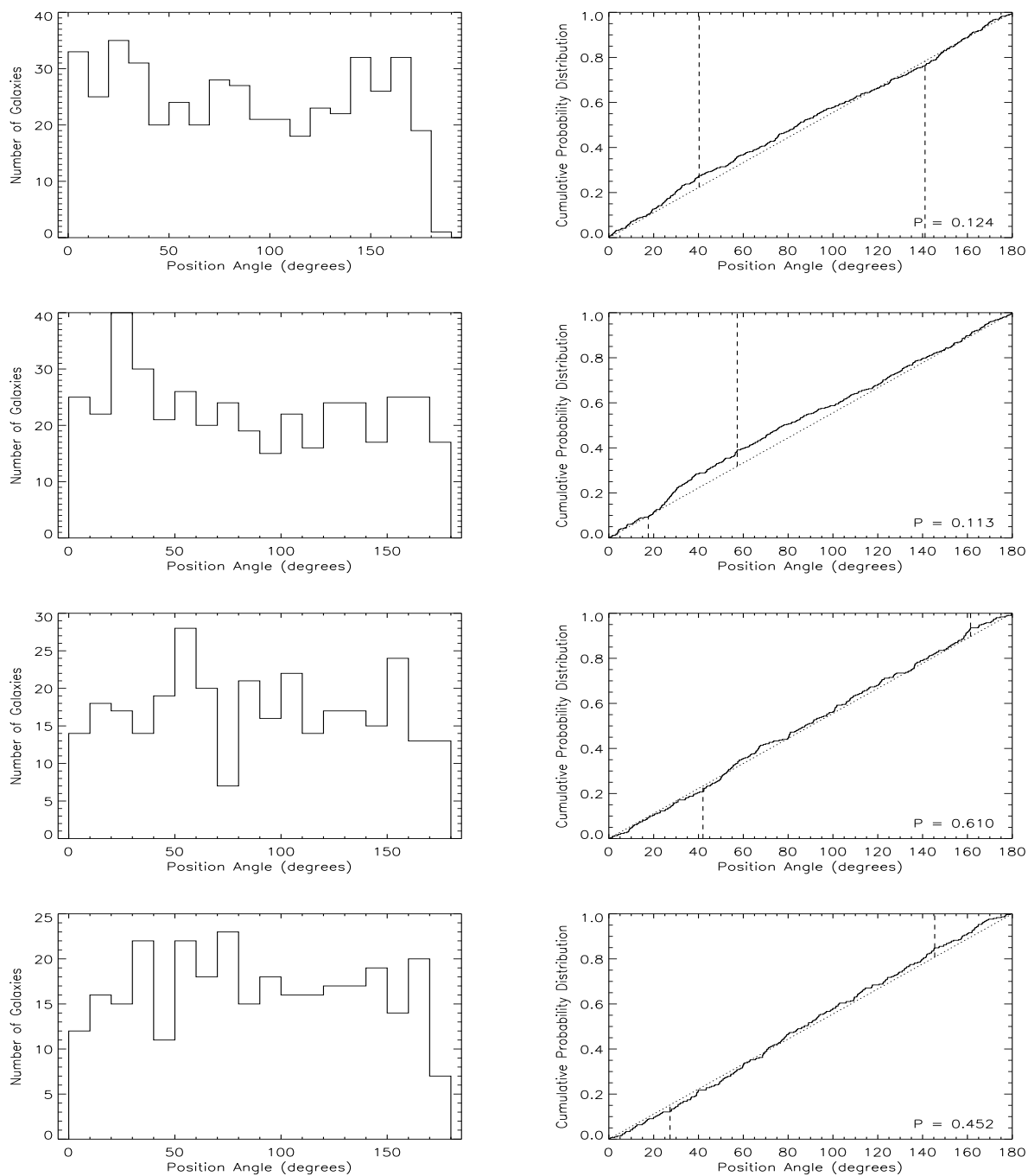}}
\caption{Histograms of position angles and probability distributions, with P value
from the Kuiper test, placed side-by-side, for galaxies in the 0 -- 200 kpc bin 
from the centre of Abell 1689 (first from top), 200 -- 400 kpc (second), 400 -- 600 
kpc (third) and 600 -- 1000 kpc (bottom). Alignment is detected for the first two 
bins (closer to the centre of the cluster) but not in the outer two.}
\end{figure}
\clearpage

\begin{figure}
\includegraphics[140,260][575,610]{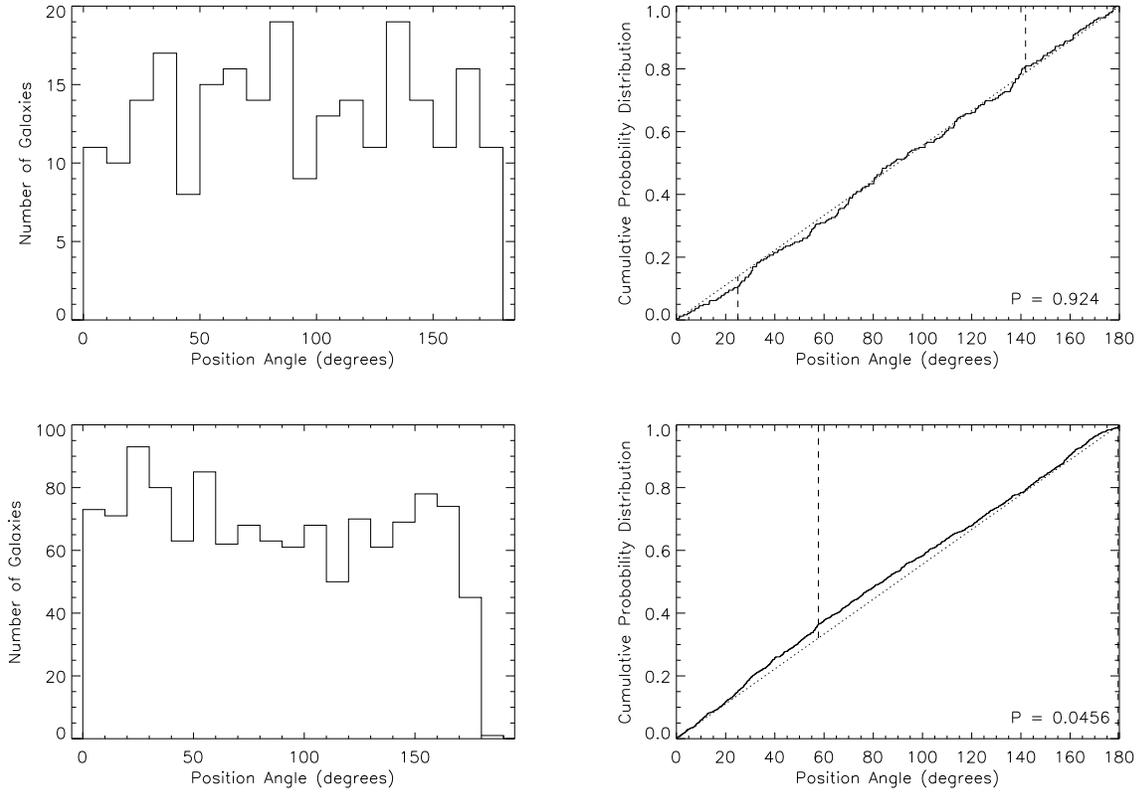}
\caption{Histograms of position angles and probability distributions, with P value
from the Kuiper test, for galaxies brighter (top panels) and fainter (bottom) than
$M_I=-18$ in Abell 1689.}
\end{figure}
\clearpage

\begin{figure}
\includegraphics[140,260][575,610]{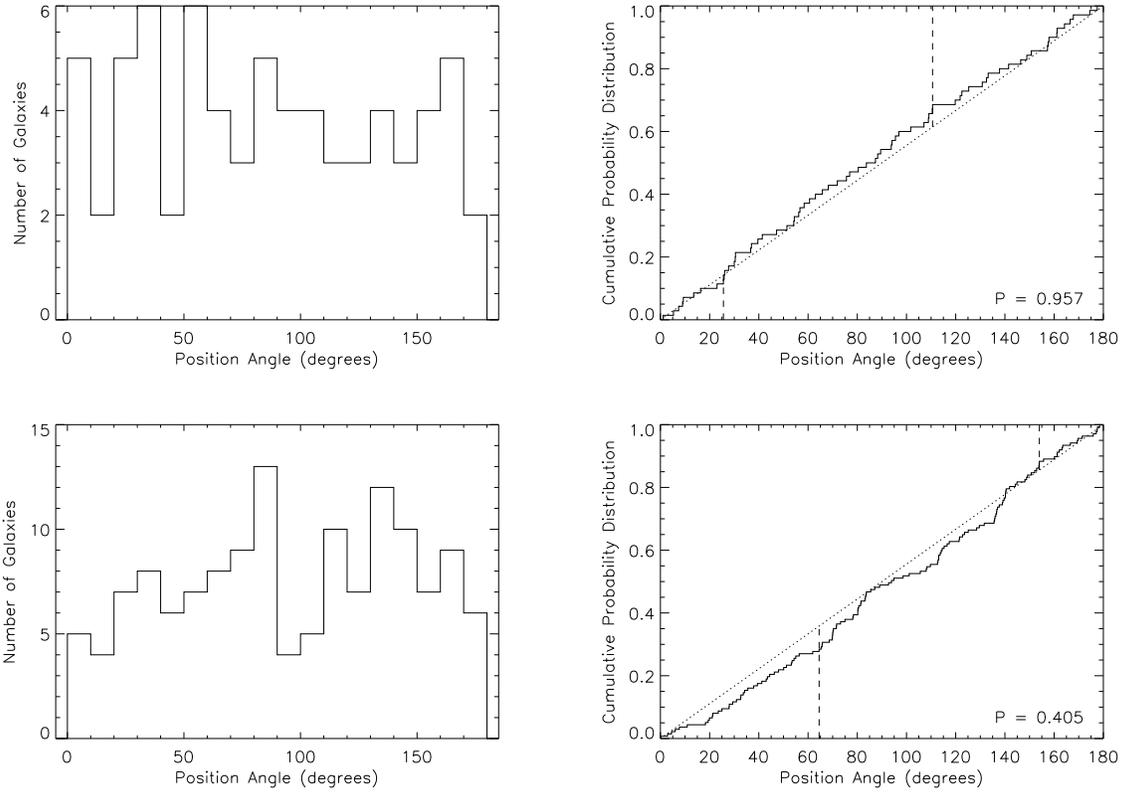}
\caption{Histograms of position angles and probability distributions, with P value
from the Kuiper test, for bright galaxies: Ellipticals in the top two panels and
S0s in the bottom two panels.}
\end{figure}

\end{document}